\documentstyle[12pt]{article}
\newcommand{\eq}{\begin{equation}}
\newcommand{\en}{\end{equation}}
\newcommand{\eqa}{\begin{eqnarray}}
\newcommand{\ena}{\end{eqnarray}}
\newcommand{\eqs}{\begin{displaymath}}
\newcommand{\ens}{\end{displaymath}}
\newcommand{\eqas}{\begin{eqnarray*}}
\newcommand{\enas}{\end{eqnarray*}}

\advance\oddsidemargin by -\textwidth
\advance\oddsidemargin by -1in
\evensidemargin=\oddsidemargin
\begin{document}

$\mbox{ }$
\vspace{-3cm}
\begin{flushright}
\begin{tabular}{l}
{\bf KEK-TH-494 }\\
{\bf KEK preprint 96 }\\

\end{tabular}
\end{flushright}
 
\baselineskip18pt
\vspace{1cm}
\begin{center}
\Large
{\baselineskip26pt \bf 
Particle-Particle-String Vertex\footnote{A talk given at Kashikojima Workshop, 
August 19- September 1, 1996.}
}
\end{center}
\vspace{1cm}
\begin{center}
\large
{ Nobuyuki Ishibashi}
\end{center}
\normalsize
\begin{center}
{\it KEK Theory Group, Tsukuba, Ibaraki 305, Japan}
\end{center}
\vspace{2cm}
\begin{center}
\normalsize
ABSTRACT
\end{center}
{\rightskip=2pc 
\leftskip=2pc 
\normalsize
We study a theory of particles interacting with strings. 
Considering such a theory for Type IIA superstring 
will give some clue about M-theory. As a first step toward such a theory, 
we construct the particle-particle-string interaction vertex generalizing 
the D-particle boundary state. 
\vglue 0.6cm}

\newpage
\section{Introduction}
Recent developments in string theory dynamics strongly suggest the existence 
of M-theory whose low energy effective theory coincides with $D=11$ 
supergravity\cite{M}. 
Knowing what M-theory really is is very important 
considering its usefulness in various aspects of string dynamics. 
M-theory might be a theory of some extended objects which yields $D=11$  
supergravity in the low energy limit. 
Many people expect that the theory of supermembranes\cite{mem} will give 
M-theory but 
quantization of supermembranes is still very difficult. 

Here we would like to take an alternative point of view. 
In order to do so, let us notice the following facts. 
M-theory can be realized 
as the strong coupling limit of Type IIA string theory\cite{witten}. 
In Type IIA superstring theory one can consider 
BPS saturated point-like objects with R-R charges, which 
can now be identified with D-particles\cite{pol}. 
The masses of these particles approach 
zero in the 
strong coupling limit and the theory looks like an eleven dimensional theory 
compactified on $S^1$ with large radius. The D-particles can be considered as 
the Kaluza-Klein modes. 

Therefore if one can construct a theory of particles (which have the same 
quantum numbers as the Kaluza-Klein modes) 
interacting with Type IIA strings like D-particles do, 
it may at least yield $D=11$ 
supergravity in the low energy limit. Since we are familiar with 
the perturbation theory of particles and strings, it is easy to quantize 
it perturbatively and see if the ultraviolet divergences cancel or not. 
Constructing such a theory will give us at least some clue about M-theory. 

In string theory, we always consider Feynmann graphs consisting of one 
kind of worldsheet which is described by a two dimensional quantum field 
theory. This is in contrast to particle theory where there are 
many kinds of particles (e.g. gauge particles, fermions, etc.) and 
the worldlines of these particles correspond to different quantum mechanics.  
Therefore it may be possible to generalize string theory so that it includes 
various kinds of worldsheets or even a theory in which various kinds of strings 
and particles 
are interacting with each other. What we would like to do is to 
construct one of the simplest of such generalizations. 

As a first step toward such a theory, we need to construct the vertex 
describing the interaction of such particles and strings. 
The most well-known interaction vertices of strings and particles are described 
by the ``vertex operators''. They represent string-string-particle interaction. 
However the particles which appear in such 
vertices are included in the spectrum of strings and these are not what we 
need. The only vertices that are known so far and describe the interaction 
of strings with something other than strings are given by the D-brane boundary 
states. They represent D-brane-D-brane-string interactions in the special 
case where D-branes are classical and at rest. The vertex we need is 
a generalization of it. 

What we will do in this note, is to construct particle-particle-string 
vertices starting from this D-brane boundary state. We here consider 
the interaction between bosonic strings and scalar particles for 
simplicity. It is straightforward to generalize it to the interaction between 
superstrings and scalar particles. The interaction between superstrings and 
higher spin particles in the same supermultiplet may be 
obtained by using the space-time supersymmetry.  

\section{Particle-Particle-String Vertex}
The first quantized action of a scalar particle is 
\eq
I=\frac{1}{2}\int dt[\frac{\dot{x}^2}{e}+em^2].
\en
Here $e$ is the einbein on the worldline and $m$ is the mass of the particle. 
This gives a scalar field with Klein-Gordon type kinetic term. 
The vertex we will construct describes interactions between this type of 
particles and strings. 

Before considering the string case, let us notice how scalar particles are 
coupled with gauge particles. 
The gauge invariant interaction between a vector particle $A_\mu$ and the 
scalar particle is expressed by the Wilson line integral 
\eq
\int dtA_\mu (x)\dot{x}^\mu, 
\label{v1}
\en
over the worldline. This is of course gauge invariant classically. 

In quantizing, one usually fixes the reparametrization invariance by putting 
$e=1$, and $\dot{x}^\mu$ becomes the momentum $p^\mu$. Therefore 
one should be careful about the ordering of the operators in eq.(\ref{v1}). 
One usually takes the ordering so that the scalar-scalar-vector vertex 
becomes $\epsilon \cdot J\equiv \epsilon \cdot (p +k/2)$, 
where $p,k$ are as in Fig. 1. 
$J_\mu=p_\mu +k_\mu /2$ can be identified with the current 
$i\phi^\dagger \partial_\mu \phi -i\partial_\mu\phi^\dagger \phi$ in the second 
quantized formalism. 
$J_\mu$ corresponds to $\dot{x}_\mu /e$ in the vertex in eq.(\ref{v1}).
\footnote{$e$ appears in this expression to make it reparametrization 
invariant. } 
If the polarization 
$\epsilon^\mu$ of the vector field is proportional to its momentum $k^\mu$, 
this vertex yields
\eq
k \cdot (p +\frac{k}{2})=\frac{1}{2}[(p+k)^2-m^2]-\frac{1}{2}(k^2-m^2).
\label{cons}
\en
This quantity vanishes if the particles involved are on-shell (i.e. 
$(p+k)^2=k^2=m^2$). If the particles are not on-shell as in the case where 
the vertex appears in a Feynmann graph, this quantity cancels the 
scalar particle propagators attached to the vertex. 
Therefore choosing an appropriate sea gull term, one can show the amplitudes 
with such a polarization vanish which means they are gauge invariant. 
Eq.(\ref{cons}) corresponds to the current conservation equation 
$\partial_\mu J^\mu =0$ in the second-quantized formalism. 

Classically, interaction with graviton $h_{\mu\nu}$ 
can be described by the vertex
\eq
\int dteh_{\mu\nu}(x)\frac{\dot{x}^\mu}{e}\frac{\dot{x}^\nu}{e} .
\label{grav}
\en
This is gauge invariant up to the equation of motion 
\eq
\partial_t(\frac{\dot{x}^\nu}{e})=0. 
\en
The quantum mechanical treatment of this vertex is much more subtle than the 
vector particle case\cite{curve}. 
The gauge variation of the quantum vertex again 
yields the inverse propagators of the particles attached to the vertex and 
can be cancelled by choosing appropriate contact interaction terms. 

Since string includes various gauge fields, what we would like to construct 
is a generalization of eq.(\ref{v1})(\ref{grav}). 
We will present it in the following form: 
\eq
\int dte|V\rangle ,
\label{int}
\en
where $|V\rangle $ is a string state. This vertex represents the couplings 
between infinitely many particles in the string spectrum with the scalar 
particle in the way that the vertex of a particle corresponding to the 
string state $|i\rangle $ is 
\eq
\int dte\langle i|V\rangle .
\en

In such a representation, the condition of gauge invariance of the vertex 
becomes as follows. The vertex of a null external state is 
\eq
\int dte\langle ~|Q_B|V\rangle .
\en
Therefore, in order for the vertex to be gauge invariant classically, 
$Q_B|V\rangle $ should vanish up to the equations of motion i.e. 
\eqa
& &
\partial_t(\frac{\dot{x}^\nu}{e})=0, 
\nonumber
\\
& &
(e)^2-\frac{\dot{x}^2}{m^2}=0.
\ena
Roughly speaking, when the scalar particle is quantized, 
these equations of motion yield 
terms which may be cancelled by choosing contact interaction terms. 
In quantizing, we should define the ordering of the operators so that 
$Q_B|V\rangle $ is written as a sum of terms proportional to the inverse 
propagators. Then by choosing appropriate contact interaction terms, the 
gauge invariance will be shown. We will not pursue these contact 
interaction terms in this note. 

Here we will first construct the classical vertex and then 
consider it quantum mechanically. 
We will construct this vertex starting from the following assumptions. 
\begin{itemize}
\item
$|V\rangle $ depends on the informations of the particles only through 
$J_\mu =\dot{x}^\mu /e$.
\item
For the particle at rest $|V\rangle $ coincides with the D-particle boundary 
state. 
\end{itemize}
The first assumption is reasonable because up to the equations of motion 
$J_\mu$ is the only independent 
quantity which is reparametrization invariant.

With the above assumptions, it is possible to obtain $|V\rangle $ uniquely up 
to the equation of motion. In order to do so, let us first rewrite the 
D-particle boundary state in terms of $J_\mu$. In that case, the particle is 
at rest and the gauge $t=x^0$ is taken. Therefore 
\eq
\vec{J}=(m,0,\cdots ,0).
\label{J}
\en
The D-particle boundary state is\cite{boundary} 
\eq
\exp [-\sum_{n>0}\frac{1}{n}\alpha_{-n}^0\tilde{\alpha}_{-n}^0
-\sum_{n>0,i}\frac{1}{n}\alpha_{-n}^i\tilde{\alpha}_{-n}^i]|k\rangle 
\otimes |B\rangle_{gh},
\en
where $|B\rangle_{gh}$ is the ghost boundary state:
\eq
|B\rangle_{gh}=\exp [-\sum_{n>0}(b_{-n}\tilde{c}_{-n}+\tilde{b}_{-n}c_{-n})]
(c_0+\tilde{c}_0)c_1\tilde{c}_1|0\rangle_{gh}.
\label{rest}
\en
$|k\rangle $ here is the Fock vacuum with momentum $k,~(k^0=0)$. We have 
made a momentum eigenstate out of the usual boundary states with Dirichlet 
boundary conditions. This boundary state is BRST invariant and represent 
the interaction vertex of the (classical) particle at rest with a string 
with momentum $k$. This boundary state can be rewritten in terms of 
$J_\mu$ in eq.(\ref{J}) as 
\eq
|V\rangle (J,k)=
\exp [-2\sum_{n>0}\frac{1}{n}
\frac{\alpha_{-n}\cdot J\tilde{\alpha}_{-n}\cdot J}{m^2}
+\sum_{n>0}\frac{1}{n}\alpha_{-n}\cdot\tilde{\alpha}_{-n}]|k\rangle 
\otimes |B\rangle_{gh}.
\label{VJk}
\en
This is the only Lorentz invariant way to do so, up to the equations of motion. 
The condition $k^0=0$ corresponds to $J\cdot k=0$. This boundary state is 
BRST invariant for any $J$ satisfying $J\cdot k=0,~J^2=m^2$. Indeed 
\eq
e|V\rangle (J,k),
\en
with $e=\sqrt{\dot{x}^2}/m$ coincides with the boundary state for the 
D-particle with constant velocity\cite{boundary}. 

Now let us construct $|V\rangle (J,k)$ with $J,k$ not necessarily satisfying 
$J\cdot k=0,~J^2=m^2$. Since $|V\rangle (J,k)$ is given as eq.(\ref{VJk}) when 
$J\cdot k=0,~J^2=m^2$, it can be expressed as\footnote{Here we assume that 
 $|V\rangle (J,k)$ depends on $J,k$ analytically. }
\eq
\exp [-2\sum_{n>0}\frac{1}{n}
\frac{\alpha_{-n}\cdot J\tilde{\alpha}_{-n}\cdot J}{m^2}
+\sum_{n>0}\frac{1}{n}\alpha_{-n}\cdot\tilde{\alpha}_{-n}]|k\rangle 
\otimes |B\rangle_{gh}
+J\cdot k|~\rangle +(J^2-m^2)|~\rangle^\prime , 
\en
using some states $|~\rangle $ and $|~\rangle^\prime$. However, since 
$J\cdot k$ and $J^2-m^2$ are proportional to the equations of motion,
\footnote{$J\cdot k$ is proportional to the equation of motion 
$\partial_t(\dot{x}^\mu /e)$ after the integration by parts in eq.(\ref{int}).}
eq.(\ref{VJk}) holds up to equations of motion for general $J,k$. 

Next thing we should do is to prove that this vertex yields a gauge invariant 
vertex. 
$|V\rangle (J,k)$ in eq.(\ref{VJk}) is BRST closed provided 
$J\cdot k=0,~J^2=m^2$. Hence $Q_B|V\rangle (J,k)$ is written as a 
sum of terms proportional to 
$J\cdot k$ and $J^2-m^2$, i.e. the equations of motion. Therefore 
$|V\rangle (J,k)$ in eq.(\ref{VJk}) gives the classical 
particle-particle-string vertex which is gauge invariant. 

In order to make the vertex gauge invariant quantum mechanically, we should 
be careful about the ordering of the operators $\dot{x}^\mu$ and $x^\mu$. 
One way to do so is to substitute $e=\sqrt{\dot{x}^2}/m$ and 
$\dot{x}_\mu=p_\mu +k_\mu /2$ into eq.(\ref{VJk})($p,k$ are as in Fig. 2):
\eq
|V\rangle (J,k)=
\exp [-2\sum_{n>0}\frac{1}{n}
\frac{\alpha_{-n}\cdot (p+k/2)\tilde{\alpha}_{-n}\cdot (p+k/2)}{(p+k/2)^2}
+\sum_{n>0}\frac{1}{n}\alpha_{-n}\cdot\tilde{\alpha}_{-n}]|k\rangle 
\otimes |B\rangle_{gh}.
\en
This makes 
$Q_B|V\rangle (J,k)$ a sum of terms proportional to the inverse propagators of 
the scalar particles attached to the vertex. 

\section{Discussions}
Now that we have constructed the interaction vertex, we can calculate various 
amplitudes. Of course, we should not forget to add 
the particle-particle-multi-string vertices to preserve the gauge symmetry. 
It is an intriguing problem to examine if such amplitudes diverges or not. 
Since the vertex in eq.(\ref{VJk}) is BRST invariant only up to the inverse 
propagators of the scalar particle, we are not sure if the string miracles 
of divergence cancellation occur in such amplitudes. Also it is possible for 
interaction vertices involving only particles to exist. If such vertices exist 
for higher spin particles, the theory becomes nonrenormalizable in general. 
Consistency of the theory will fix if such vertices are necessary or not. 
All these problems will be studied elsewhere. 

\section*{Acknowledgements}
We would like to thank M. Bando, H. Ishikawa, H. Kawai, Y. Kazama, H. Kunitomo, 
Y. Matsuo and M. Natsuume for useful discussions and comments. We are grateful 
to the organizers of Kashikojima workshop for organizing such a enjoyable and 
stimulating workshop.

\newpage
\section*{Figure Captions}
\begin{description}
\item{Fig. 1} Scalar-scalar-vector vertex.  
\item{Fig. 2} Scalar-scalar-string vertex. 
\end{description}
\end{document}